\documentclass[pre,superscriptaddress,showpacs,aps]{revtex4}
\usepackage{graphicx}
\usepackage{dcolumn}
\usepackage{bm}
\usepackage{amsmath}
\usepackage{color}
\usepackage{epsfig}

\begin{document}


\title{Ultrarelativistc regime in the propagation of an ultrastrong,\\ femtosecond laser pulse in plasmas}

\author{Du\v san Jovanovi\'c}
\email{djovanov@ipb.ac.rs} \affiliation{Institute of Physics, University of Belgrade, Pregrevica 118,
11080 Belgrade (Zemun), Serbia}
\author{Renato Fedele}
\email{renato.fedele@na.infn.it} \affiliation{Dipartimento di Fisica, Universit\`{a} di Napoli "Federico II", 
M.S. Angelo,
Napoli, Italy}
\affiliation{INFN Sezione di Napoli, Complesso Universitario di M.S. Angelo, Napoli, Italy}
\author{Milivoj Beli\'c}
\email{milivoj.belic@qatar.tamu.edu} \affiliation{Texas A\&M University at Qatar, P.O. Box 23874 Doha, Qatar}
\author{Sergio De Nicola}
\affiliation{SPIN-CNR, Complesso Universitario di M.S. Angelo, Napoli, Italy}
\affiliation{Dipartimento di Fisica, Universit\`{a} di Napoli "Federico II", 
M.S. Angelo,
Napoli, Italy}
\date{\today}

\begin{abstract}

The interaction of a multi-Petawatt, pancake-shaped laser pulse with an unmagnetized plasma is studied analytically and numerically in the regime of fully relativistic electron jitter velocities and in the context of the laser wakefield acceleration scheme. The study is applied to the specifications available at present time, or planned for the near future, of the Ti:Sa Frascati Laser for Acceleration and Multidisciplinary Experiments (FLAME) in Frascati. A set of novel nonlinear equations is derived using a three-timescale description, with an intermediate timescale associated with the nonlinear phase of the electromagnetic wave and with the spatial bending of its wave front. They describe on an equal footing both the strong and moderate laser intensity regimes, pertinent to the core and the edges of the pulse. These have fundamentally different dispersive properties since, in the core, the electrons are almost completely expelled by a very strong ponderomotive force and the electromagnetic wave packet is imbedded in a vacuum channel and has (almost) linear properties, while at the pulse edges the laser amplitude is smaller and the wave is dispersive. The new nonlinear terms in the wave equation, introduced by the nonlinear phase, describe a smooth transition to a nondispersive electromagnetic wave at very large intensities, and the simultaneous saturation of the previously known nonlocal cubic nonlinearity, without the violation of the imposed scaling laws. The temporal evolution of the laser pulse is studied by the numerical solution of the model equations in a two-dimensional geometry, with the spot diameter presently used in the self-injection test experiment (SITE) with FLAME. The most stable initial pulse length is found to be around 1 $\mu$m, which is several times shorter than presently available. A rapid stretching of the laser pulse in the direction of propagation is observed, followed by the development of a vacuum channel and a very large electrostatic wake potential, as well as the bending of the laser wave front.

\end{abstract}

\maketitle

\section{Introduction}\label{Introductory}
A theoretical investigation of the interaction of an ultra-strong and ultra-short laser pulse with unmagnetized plasma is carried out, aimed for the advancement of the laser wakefield acceleration scheme. The study is applied to the specifications available at present time, or planned for the near future, of the Ti:Sa Frascati Laser for Acceleration and Multidisciplinary Experiments (FLAME), as well as to the characteristics envisaged for the next generation of powerful lasers. The analysis is based on the Lorentz-Maxwell fluid model in the fully relativistic regime taking the pancake approximation, developed in earlier publications. In our recent paper \cite{moj_EPJD} we studied in detail only the WIR (Weak Intensity Regime) and MIR (Moderate Intensity Regime), defined below, at the end of section \ref{model}. The SIR (Strong Intensity Regime) was discussed in \cite{moj_EPJD} only qualitatively and it was pointed out that it was fundamentally different from MIR, since it involves vastly different scalings in the core and at the edges of such pulse. Namely, the electrons are almost completely expelled from the core by a very strong ponderomotive force, creating a {\em vacuum channel}. An electromagnetic wave packet is imbedded in such vacuum channel and features (quasi)linear properties, as if it was propagating virtually in vacuum. Conversely, the edges of the pulse (and most importantly, the leading edge) operate within the MIR, and the sort on nonlinear self-organization described in Ref. \cite{moj_EPJD} is expected to occur there. Thus, in order to study the propagation of a very large amplitude pulse, we need a general description that includes both the (quasi)linear bahavior inside the vacuum channel and the proper boundary conditions at its edges, including the creation of such vacuum channel by the electron expulsion at the leading edge of the pulse.

Plasmon-X is a facility located in the Frascati INFN laboratories based on the Ti:Sa laser FLAME and electrons' linac SPARC. The characteristics of the FLAME laser, that is used in the Plasmon-X device, are \cite{Prometeus} $E= 7$ J, $\tau\geq$25 fs, $W\leq 300$ TW, $\lambda = 0.8\, \mu$m, (which corresponds to $\omega = 2.35619\times 10^{15}\,{\rm s}^{-1})$, and $\nu_{rep}= 10$ Hz (repetition frequency). It is worth noting that the pulse duration $T = 25\times 10^{-15}\,{\rm s}$ corresponds to the pulse length $L_z = 7.5\,\,\mu{\rm m}$, i.e. there are around 10 wavelengths of the laser light within the pulse. An upgrade that includes the polarization control (S, P, circular) is planned for the near future \cite{Prometeus}. The plasma density in different Plasmon-X experiments ranges as $n_e = 0.6-1\times 10^{19}\, {\rm cm}^{-3}$, and up to $4\times 10^{19}\, {\rm cm}^{-3}$ \cite{NTA,Gizzi}. The electron density $n_e = 10^{19}\, {\rm cm}^{-3}$ corresponds to the plasma frequency $\omega_{p,e} = 1.784\times 10^{14}\,s^{-1}$. Thus, the pulse duration is roughly $\tau = 0.7 \,\, T_p$, where $T_p$ is the plasma period $T_p \equiv 2 \pi/\omega_{p,e} = 35.22\times 10^{-15}\,{\rm s}$, while the collisionless skin depth (which is the wavelength of the natural oscillation mode of the plasma), $d_e \equiv 2\pi c/\omega_{p,e} = 10.57\,\,\mu$m is an order of magnitude longer than the laser wavelength ($\lambda = 0.8\,\, \mu$m) and close to the pulse length ($L_z = 7.5\,\,\mu{\rm m}$). The other parameters, such as the laser spot size and the plasma length, have different values in various experiments for the laser wakefield acceleration of electrons (including numerical experiments). In the simulations of the external injection of electrons \cite{NTA}, the plasma was taken to be 9.88 cm long with a density profile with a positive and varying slope, whose starting and ending densities were $1.5\times 10^{17} \, {\rm cm}^{-3}$ and $2.5\times 10^{17} \, {\rm cm}^{-3}$, respectively, while the laser pulse was taken to have the initial waist size of $130\, \mu$m and minimum size of $32.5\, \mu$m, guided by a matched channel profile. Considerably smaller plasma lengths were used for the electron acceleration at the sub GeV energy level in the self-injection test experiment (SITE) using FLAME \cite{Gizzi2013,GizziNuovoCim}. In that laser wakefield accelerator, that will be used for the excitation of an all-optical X-ray radiation source at LNF (Laboratori Nazionali di Frascati), two different gas-jet configurations are utilized with helium gas at 25 bar, and the nozzle being a 4 mm long, 1.2 mm wide slit. Correspondingly, the plasma lengths inside the gas jet was 10 mm (longitudinal propagation) and 4 mm or 1.2 mm (transverse propagation).

In the present paper, we consider the SIR laser intensities of the order of $I \sim 10^{20} \, {\rm W/cm}^2$, which is 30-50 times bigger than that realized in FLAME laser-wakefield experiments. In the FLAME, such intensities are envisaged only for the Thomson scattering scheme, for the production of the intense pulses of X (or $\gamma$) radiation, where they are realized by a strong laser focusing to a spot size $\lesssim 10 \, \mu{\rm m}$. The latter is not convenient for the acceleration scheme, but such laser intensities will be much more efficient for the particle acceleration, and they are expected to be reached with the next generation lasers even with a $\sim 100 \, \mu{\rm m}$ spot. This can be done by increasing the laser energy 10--20 times, to $E \approx 125 \, {\rm J}$, and decreasing the pulse length 3--5 times, which yields a laser power of several tens of Petawatt. In order to make predictions for the behavior of such strong pulses, we derive a novel mathematical model that describes both the moderate and the strong intensity regimes. In the classical picture of a slowly varying amplitude of the laser pulse, based on a two-timescale description, this is not possible because the dispersion characteristics of electromagnetic waves in MIR and SIR are too different from each other and can not be described on a common footing. In the core of a very strong (i.e. SIR) pulse, the electromagnetic wave practically propagates in a vacuum. Such wave is not dispersive, i.e. its group velocity is constant and coincides with its phase velocity. Conversely, at the edges of such pulse the amplitude is smaller and the wave is dispersive. Under such conditions, the simple envelope description used previously in the MIR, breaks down. Our model is derived using a three-timescale description, with an intermediate timescale associated with the nonlinear, intensity-dependent, phase of the electromagnetic pulse. The Schr\"{o}dinger equation for the phase is considerably simplified under the physical conditions of the FLAME laser system (such as the laser frequency, pulse duration and spot size, plasma density etc.). For the laser power that is around 170 times bigger than that used at present time in the Plasmon-X laser wakefield experiments in Frascati, our equation for the phase can be solved within the WKB (Wentzel–Kramers–Brillouin) approximation. The new nonlinear terms in the wave equation, introduced by the nonlinear phase, describe a smooth transition to a nondispersive electromagnetic wave at very large intensities, and the simultaneous saturation of the previously known nonlocal cubic nonlinearity, without the violation of the imposed scaling laws. These equations are solved numerically in a 2-D (two-dimensional) geometry. A violent stretching of the laser pulse in the direction of propagation is observed, which permits the pulse to propagate through plasma up to a several mm distance, which is consistent with the conditions of the self-injection experiment \cite{Gizzi2013,GizziNuovoCim}. The stretching is attributed to the nonlocality effects, which give rise to an effective mixing of the core of the pulse (which propagates with the speed of light through a self-generated vacuum channel) with the front edge of the pulse (which tends to propagate with the group velocity).

\section{Mathematical model}\label{model}

We do not include here the derivation of the coupled system (wave equation + Poisson's equation) that involves fully relativistic electrons. It is based on the classical works \cite{mahajan,Shukla,sharma_ordinary_NLS,Vulcan} and the details of its derivation are given also in our recent paper \cite{moj_EPJD}. These equations are valid in an unmagnetized plasma, provided the solution is slowly varying in the reference frame that moves with the velocity $u\,\vec{e}_z$, and they are written in the following dimensionless quantities
\begin{equation}\label{dimensless}
\vec{p'} = \frac{\vec p}{m_0 c},\quad \vec{v'} = \frac{\vec
v}{c},\quad \phi' = \frac{q\phi}{m_0 c^2},\quad \vec{A'} =
\frac{q\vec A}{m_0 c}, \quad n'= \frac{n}{n_0}, \quad
u'=\frac{u}{c},
\quad
t' = \omega_{pe} t, \quad {\vec r\,\,}' =
\frac{\omega_{pe}}{c}\left( \vec r - \vec e_z\, u t\right),
\end{equation}
where $\omega_{pe}$ is the electron plasma frequency of of the unperturbed plasma, $\omega_{pe} = (n_0 q^2/m_0 \epsilon_0)^\frac{1}{2}$, while $-e$ and $m_0$ are the electron charge and the rest mass. For simplicity, the primes will be omitted in the rest of the paper. The components of the Maxwell's equations that are parallel and perpendicular to the direction of the electromagnetic wave propagation (i.e. the wave equation for the perpendicular component of the vector potential, $\vec A_\bot$, and the Poisson's equation for the electrostatic potential $\phi$) take the form
\begin{eqnarray}
\label{bdnorwav} && \left[\frac{\partial^2}{\partial t^2} -2 u
\,\frac{\partial^2}{\partial z \, \partial t} - \left(1 -
u^2\right)\frac{\partial^2}{\partial z^2}-\nabla_\bot^2\right]
\vec A_\bot + \nabla_\bot\left(\frac{\partial}{\partial t}-u\,
\frac{\partial}{\partial z}\right)\phi
= \vec v_\bot n \\
\label{bdPoisson} && \left(\nabla_\bot^2 + \frac{\partial^2}{\partial z^2}\right)\phi = 1-n
\end{eqnarray}
while the electron continuity, the longitudinal and  the perpendicular component of the momentum equations are
\begin{eqnarray}
\label{bdcont} && \left(\frac{\partial}{\partial t} - u\,\frac{\partial}{\partial z}\right) n + \nabla\cdot\left(n\vec v\right) = 0 \\
\label{bdparmom} && \left(\frac{\partial}{\partial t} - u\,
\frac{\partial}{\partial z} + \vec
v_\bot\cdot\nabla_\bot\right)\left(p_z + A_z\right) - \vec
v_\bot\,\frac{\partial}{\partial z}\left(\vec p_\bot + \vec
A_\bot\right) + \frac{\partial}{\partial z}\left(\gamma +
\phi\right) = 0 \\
\label{bdnormom} && \left[\frac{\partial}{\partial t} + \left(v_z
- u\right)\frac{\partial}{\partial z} + \vec v_\bot \cdot
\nabla_\bot \right] \left(\vec p_\bot + \vec A_\bot\right) -
v_i\nabla_\bot\left(p_i + A_i\right) + \nabla_\bot\left(\gamma +
\phi\right) = 0,
\end{eqnarray}
where $\gamma$ is the relativistic factor, $\gamma = (1 + \vec p^{\, 2}/m_0^2 c^2)^\frac{1}{2}$,
and $c$ is the speed of light.

As already mentioned, the solution of the hydrodynamic equations (\ref{bdcont}), (\ref{bdparmom}), and (\ref{bdnormom}) is sought in a {\em quasistatic regime}, i.e. when the solution is slowly varying in the moving reference frame, viz.
$
{\partial}/{\partial t}\ll u \, {\partial}/{\partial z}
$. 
However, the hydrodynamic equations (\ref{bdcont})-(\ref{bdnormom}) still remain rather complicated, and following the classical works \cite{mahajan,Shukla,sharma_ordinary_NLS,Vulcan}, we further simplify them by adopting $u$ to be very close to the speed of light
$
1-u\ll 1
$. 
In Ref. \cite{moj_EPJD} we generalized the results of \cite{mahajan,Shukla,sharma_ordinary_NLS,Vulcan} to a 3-D geometry, but assuming a {\em pancake} (i.e. almost one dimensional, 1-D) solution, viz.
$
\nabla_\bot \ll {\partial}/{\partial z},
$. 
In the approximate expressions for the charge and current densities, we use the leading order solution of the electron hydrodynamic equations (\ref{bdcont})-(\ref{bdnormom}), which is found as a stationary 1-D solution that is propagating with the speed of light, setting $\partial/\partial t = \nabla_\bot = 1-u = 0$. Then, the leading parts of Eqs. (\ref{bdcont}), (\ref{bdparmom}), and (\ref{bdnormom}) are obtained in a simple form
\begin{eqnarray}
\label{bdcontL}   &&  \frac{\partial}{\partial z}\left[\left(v_z - 1\right)n\right] = 0, \\
\label{bdnormomL} &&  \frac{\partial}{\partial z}\left(- p_z +\gamma + \phi\right) = 0, \\
\label{bdparmomL} &&  \frac{\partial}{\partial z}\left(\vec p_\bot
+ \vec A_\bot\right) = 0,
\end{eqnarray}
while from $\nabla\cdot\vec A = 0$, within the same accuracy, we
have
\begin{equation}\label{parAL}
{\partial A_z}/{\partial z} = 0.
\end{equation}
Noting that for $z\to\pm\infty$ we have $\phi = \vec A = \vec v =
\vec p = 0$ and $\gamma = n = 1$, and using $\gamma = (1 + p_z^2 +
{\vec p_\bot}^{\,\,2})^\frac{1}{2}$, Eqs.
(\ref{bdcontL})-(\ref{parAL}) are readily integrated, yielding
\begin{eqnarray}
\label{bdcontL2}   &&  \left(v_z - 1\right)n + 1 = 0, \\
\label{bdnormomL2} &&  - p_z +\gamma -1 + \phi = 0, \\
\label{bdparmomL2} &&  \vec p_\bot + \vec A_\bot = 0,\\
\label{parAL2} &&  A_z = 0.
\end{eqnarray}
Then, making use of Eqs. (\ref{bdcontL2})-(\ref{bdparmomL2}) and the definition of $\gamma$, we obtain dimensionless charge and current densities as
\begin{eqnarray}
\label{n2}  &&   n = \frac{\left(\phi - 1\right)^2 +
{\vec{A}_\bot}^{\, 2} + 1}{2 \left(\phi - 1\right)^2}, \\
\label{jn2}  && \vec v_\bot n = \frac{\vec A_\bot}{\phi - 1},
\end{eqnarray}
which permits us to rewrite our basic equations as
\begin{equation}\label{waveqenv0}
\left[\frac{\partial^2}{\partial t^2} - 2
u\,\,\frac{\partial^2}{\partial t\,\, \partial z} -
\left(1-u^2\right)\frac{\partial^2}{\partial z^2} - \nabla_\bot^2
+ \frac{1}{1-\phi}\right]\vec A_\bot =
-\left(\frac{\partial}{\partial t} - u\,\,
\frac{\partial}{\partial z}\right)\nabla_\bot\phi,
\end{equation}
\begin{equation}\label{poissons0}
\frac{\partial^2\phi}{\partial z^2} = \frac{\left(\phi -
1\right)^2 - 1 - {\vec{A}_\bot}^{\, 2}}{2 \left(\phi - 1\right)^2},
\end{equation}
The above equations (\ref{waveqenv0}) and (\ref{poissons0}) constitute a system of coupled nonlinear
equations that describe the spatio-temporal evolution of an electromagnetic wave, modulated in the form of a pancake,
interacting with a Langmuire wave via the nonlocal nonlinearities
that arise from the relativistic effects, beyond the slowly
varying amplitude approximation and for an arbitrary intensity
regime. They appropriately describe all the parametric
processes involved, i.e. besides the standard nonrelativistic three-wave
coupling (the Raman scattering), they provide also the description of the
four-wave processes in a relativistic plasma, directly related to the modulational
instability, the soliton formation, etc.

The scaling analysis of the equations (\ref{waveqenv0}) and (\ref{poissons0}) indicates that three fundamentally different regimes of operation can be distinguished with respect to the magnitude of the nonlinear term. These are the \textit{weak intensity regime} (WIR), $\vec{A}_\bot^{\, 2} \ll \epsilon^2 \ll 1$, \textit{moderate intensity regime} (MIR), $\vec{A}_\bot^{\, 2} \sim \epsilon^2$, and \textit{strong intensity regime} (SIR), $\vec{A}_\bot^{\, 2} \sim 1$. The small parameter $\epsilon$ is defined as $\epsilon = \omega_{pe}/\omega$, which under the conditions of the FLAME laser wakefield experiments has the value $\epsilon\approx 1/12$. The dimensionless laser intensity $\vec{A}_\bot^{\, 2}$ is expressed via the laser parameters as
\begin{equation}\label{intensity}
\vec{A}_\bot^{\, 2} =
\frac{I}{c^2\epsilon_0}\frac{\lambda^2}{4\pi^2 c}\left(\frac{e}{m_0 c}\right)^2.
\end{equation}
Here $\lambda$ and $I$ are the laser wavelength and the intensity, $I = 4 W/ L_\bot^2 \pi$, while $W$ is the laser power and $L_\bot$ is the diameter of the spot. With the presently available maximum FLAME laser power of $W\sim 220$ TW and pulse duration $T \sim 25\times 10^{-15}\,{\rm s}$, all these intensity regimes have been envisaged in different experiments proposed in the literature. The corresponding laser intensities are:

\noindent (i) \textit{weak intensity regime} (WIR), with the maximum intensity $I_{max}$ ranging from $2.5\times 10^{14}$ to $2.5\times 10^{16}$ W/{\rm
cm}$^2$\,\cite{Conceptual_Design_Report,PLASMONXproject,2006PhPl...13i3103G};

\noindent (ii) \textit{moderate intensity regime} (MIR), with $I_{max}$ ranging from $1.5\times 10^{18}$ to $3\times 10^{19}$ W/{\rm cm}$^2$ \,\cite{Broggi,2006PhPl...13i3103G};

\noindent (iii) \textit{strong intensity regime} (SIR), with $I_{max}$ ranging from $10^{20}$ to $2.5\times 10^{22}$ W/{\rm cm}$^2$\, \cite{NTA,Gizzi,2006PhPl...13i3103G}.

In accordance with the FLAME specifications, the strong intensity regime might be accessed by means of the phase front correction, planned in the near future, and using very fast focusing optics, with which a 1$\mu$m diameter focal spot can be reached. However, the configurations with such small spot size provide only a short interaction length with the plasma, and for that reason they are not used in the accelerator scheme that is of our primary interest here. They will be used mostly in the Thomson scattering scheme for the production of the intense pulses of X (or $\gamma$) rays. It should be noted also that for such strong focussing, the longitudinal and transverse scales of the pulse are comparable, and the simple 1D approximation, Eqs. (\ref{bdcontL}) - (\ref{parAL}) is not applicable. Such fully 3D regime featuring very large perturbations of the electron density is often referred to as the {\em bubble regime}. Due to its inherent complexity, at present time no reliable analytic analyses of the laser-plasma interaction in the bubble regime have been presented in the literature, and all studies relied solely on the massive PIC numerical simulations. However, in the near future, a regime with a similar laser intensity, for which an analytical or semi analytic treatment will be possible, is expected be realized also in the 'pancake' geometry. With the presently available laser energies, a strong intensity pancake pulses would require a simultaneous reduction of both the spot diameter and the pulse length by the modest factor of $\sim 3$. The latter might be achieved using various techniques, such as the transition through periodic plasma-vacuum structures and plasma lenses \cite{pukhov-compress,KatsouleasPlasmaLens}, self-focusing by the plasma wake \cite{Katsouleas-compress,balakin-compress,Pipahl-compress} or  non-uniform Ohmic heating by the laser pulse, see \cite{Sharma-compress} and references therein. Together with a slight reduction of the spot diameter, readily available at present time, this would bring us to the verge of the SIR, but still retaining the pancake form of the pulse.
Conversely, it is reasonable to expect that in the foreseeable future, the laser energy can be increased 50--100 times, to the level that enables experiments in a strong intensity regime, even with the presently available pulse lengths.

\section{Wave modulation and the scaling laws}\label{scaling}

\subsection{Beyond the standard modulational representation. Nonlinear phase evolution on an intermediate scale}

For both moderate and strong intensities, we seek the solution of the wave equation in the moving frame (\ref{bdnorwav}) as the sum of a slowly varying component and a modulated electromagnetic wave, including a phase $\varphi$ that is varying on an intermediate scale, viz.
\begin{equation}\label{modwav}
\vec A_\bot = \vec A^{(0)}_\bot\left(t_2,\vec{r}_2\right) +
\left\{ \vec A_{\bot_0}\left(t_2,\vec{r}_2\right) \,\, e^{i\left[\varphi\left(t_1,\vec{r}_1\right) - \omega' t + k'\left(z + ut \right)\right]} + c.c.\right\}
\end{equation}
The slowly varying vector potential $\vec A^{(0)}_\bot\left(t_2,\vec{r}_2\right)$ corresponds to the self-generated quasistationary magnetic field. The dimensionless frequency $\omega'$ and the dimensionless wavenumber $k'$ of the rapidly varying part of the vector potential are defined as
\begin{equation}\label{nnmfrekwn}
\omega'=\frac{\omega}{\omega_{pe}}, \quad k' =
\frac{ck}{\omega_{pe}} = \frac{d_e}{\lambda},
\end{equation}
while $\omega$, $k$, and $\lambda$ are, respectively, the frequency, the wavenumber, and the wavelength of the electromagnetic wave propagating in an unperturbed plasma, that satisfy the linear dispersion relation $\omega = \sqrt{c^2 k^2 + \omega_{pe}^2}$. For simplicity, hereafter, we drop the primes and write the dimensionless version of the dispersion relation as
\begin{equation}\label{lindisprel}
\omega = \sqrt{k^2 + 1} .
\end{equation}
The quantity $\epsilon \equiv 1/\omega \ll 1$ is a small parameter under the conditions of the FLAME laser-plasma experiments, for which we have $\omega\approx k \gtrsim 12$. Likewise, the spatial derivative in the direction of propagation is estimated as ${\partial\vec A_{\bot_0}}/{\partial z}\sim {\vec A_{\bot_0}}/{L_\Vert}$, where $L_\Vert = T c = 7.5\,\mu{\rm m}$ is the pulse length and $T$ is the pulse duration. The dimensionless value of the pulse length, see Eq. (\ref{dimensless}), is then $L_\Vert' = T \omega_{pe} = 4.46$.
The intermediate and slow timescales, introduced in Eq. (\ref{modwav}) and denoted by the subscripts $1$ and $2$, respectively, are given by
\[
t_1 = \epsilon\, t - \epsilon^{-1} u z, \quad
\vec{r}_1 = \vec{e}_x x + \vec{e}_y y + \epsilon^{-1}\vec{e}_z\, z, \quad
\]
\begin{equation}\label{othernot}
t_2  = \epsilon^2 t_1= \epsilon^3 t - \epsilon\, u z, \quad
\vec{r}_2 = \epsilon\, \vec{r}_1 = \epsilon \left(\vec e_x x + \vec e_y y\right) + \vec e_z z,
\end{equation}
while the nonlinear phase $\varphi(t_1,\vec{r}_1)$ will be conveniently adopted later. We adopt $u$ to be equal to the group velocity of an electromagnetic wave
\begin{equation}\label{groupvel}
u = \frac{d\omega}{d k} = \frac{k}{\omega},
\end{equation}
which permits us to rewrite the wave equation (\ref{waveqenv0}) and the Poisson's equation (\ref{poissons0}) as
\[
2\, {\rm Re}\left\{e^{i\left[\varphi\left(t_1,\vec{r}_1\right)-t/\omega+k z\right]}\left[\alpha\, \vec{A}_{\bot_0} -
2\, i\, \epsilon^2 \left(1-\frac{\partial\varphi}{\partial t_1}\right)\frac{\partial \vec{A}_{\bot_0}}{\partial t_2} - 2\,i \, \epsilon \left(\nabla_1\varphi\cdot\nabla_2\right) \vec{A}_{\bot_0}  + \epsilon^4\,\frac{\partial^2 \vec{A}_{\bot_0}}{\partial t_2^2} - \epsilon^2\,\nabla_2^2 \vec{A}_{\bot_0}
\right]\right\} =
\]
\begin{equation}
\label{waveqenv012}
\epsilon\left(\epsilon\,\frac{\partial}{\partial t_2}- u \, \frac{\partial}{\partial z_2}\right){\nabla_2}_\bot\phi -
\left(\epsilon^4\,\frac{\partial^2}{\partial t_2^2} - \epsilon^2\nabla_2^2 - \frac{1}{1-\phi}\right)\vec{A}^{(0)}_\bot,
\end{equation}
\begin{equation}\label{poissons1}
\left(\frac{\partial}{\partial z_2} - \epsilon\, u \, \frac{\partial}{\partial t_2}\right)^2\phi = \frac{\left(\phi -
1\right)^2 - 1 - {\vec{A}_\bot}^{\, 2}}{2 \left(\phi - 1\right)^2},
\end{equation}
where
\begin{equation}\label{defalfa}
\alpha = \left(\nabla_1\varphi\right)^2 - i \, \nabla_1^2\varphi + 2\frac{\partial\varphi}{\partial t_1} - \left(\frac{\partial\varphi}{\partial t_1}\right)^2 + i \, \frac{\partial^2\varphi}{\partial t_1^2}+ \frac{\phi}{1-\phi},
\end{equation}
and $\nabla_k = \vec{e}_x (\partial/\partial x_k) + \vec{e}_y (\partial/\partial y_k) +\vec{e}_z (\partial/\partial z_k)$, $k=1,2$. The right-hand-side of the wave equation (\ref{waveqenv012}) is slowly varying in space and time and, for that reason, it is not resonant with the high-frequency oscillations of the vector potential on the left-hand-side. The slow component of the vector potential, $A_\bot^{(0)}$, is related with the quasi stationary magnetic field generated in the laser-plasma interaction, for whose accurate description it would be necessary to include also the kinetic effect that are responsible e.g. for the off-diagonal terms in the stress tensor for electrons \cite{Stress-magn-polje,Ja-i-Boba}, for the return electron current \cite{Ruski-rivju,Pegoraro-i-Califano,Askar-an}, etc. For the structure of the magnetic field created in the Weibel instability of the return current see e.g. \cite{Ja-i-Pegoraro} and references therein. However, to be consistent with the derivation used in this paper which is based on a cold and unmagnetized plasma model, we have to restrict our analysis to the regime when the right-hand side of Eq. (\ref{waveqenv012}) is negligible.

\subsection{Nonlinear phase}

First, we adopt the phase $\varphi$ to be a nonlinear function of the wake potential $\phi$, satisfying the equation
\begin{equation}\label{eqvarphi}
\alpha - \frac{\phi}{1-\phi} + i \left(\nabla_1^2\varphi - \frac{\partial^2\varphi}{\partial t_1^2}\right) = \left(\nabla_1\varphi\right)^2 + 2\,\frac{\partial\varphi}{\partial t_1} - \left(\frac{\partial\varphi}{\partial t_1}\right)^2 = \kappa^2\left(\phi\right),
\end{equation}
where $\kappa(\phi)$ is a localized, well-behaved function of its argument, that will be conveniently adopted later. With such choice of $\kappa$, the right-hand-side of Eq. (\ref{eqvarphi}) is varying on the same temporal and spatial scales as the wake potential $\phi(t_2, \vec{r}_2)$, see Eq.s  (\ref{poissons0}) and (\ref{poissons1}), i.e. we have the scaling $\kappa^2(\phi) \sim(t_2,  \vec{r}_2) = (\epsilon^2 t_1, \epsilon\, \vec{r}_1)$. In other words, the function  $\kappa^2(\phi)$ is adopted to be a slowly varying function of the spatial variables $\vec{r}_1$ and a {\em very slowly} varying function of the temporal variable $t_1$. With such choice, the fundamental solution of Eq. (\ref{eqvarphi}), which we will use as the nonlinear phase $\varphi$ in the rest of the paper, is a slowly time-varying function, viz. $\partial\varphi/\partial t_1 \ll 1$.

To be consistent with the stationary, 1-D approximation used in the derivation of the Poisson's equation (\ref{poissons0}), we need to solve Eq. (\ref{eqvarphi}) with the accuracy to $\epsilon^2$. First, we note that it can be considerably simplified if we introduce the following new variables
\begin{equation}\label{rettime1}
\vec{\rho} = \vec{r}_1, \quad\quad
\tau = t_1 - \int^{\vec{r}_1}_{-\infty} \frac{\vec{dl}\cdot\nabla_1\varphi}{\left(\nabla_1\varphi\right)^2},
\end{equation}
yielding
\begin{equation}\label{newde3riv}
\frac{\partial}{\partial t_1} = \left[1 - \frac{\partial}{\partial t_1}\int^{\vec{r}_1}_{-\infty}      \frac{\vec{dl}\cdot\nabla_1\varphi}{\left(\nabla_1\varphi\right)^2}\right] \frac{\partial}{\partial \tau}, \quad\quad
\nabla_1 = \nabla_\rho - \frac{\nabla_1\varphi}{\left(\nabla_1\varphi\right)^2}\, \frac{\partial}{\partial \tau}.
\end{equation}
Now, Eq. (\ref{eqvarphi}) takes the simple form
\begin{equation}\label{eqvarphi2}
\left(\nabla_\rho\varphi\right)^2 - \kappa^2\left(\phi\right) = {\cal O}\left(\epsilon^4\right),
\end{equation}
where we used the notation $\nabla_\rho  = \vec{e}_j \,(\partial/\partial \rho_j)$. In Eq. (\ref{eqvarphi2}), the derivatives with respect to the retarded time $\tau$ appear only in the small terms, of order ${\cal O}( \epsilon^4)$, and can be neglected. We also note that $\kappa(\phi)$ (which quantitatively describes the nonlinear laser--plasma interaction) is a slowly varying function of the spatial variable $\vec\rho$, since $\phi = \phi(\epsilon^2\tau, \, \epsilon \, \vec{\rho})$. From a physical reasoning (we will demonstrate this point later), we also intuitively expect that $\kappa^2$ is a positive definite function, i.e. that $\kappa$ does not have any zeros on the $\vec\rho$ plane. Rather, it exponentially tends to zero for $|\vec\rho|\to\infty$, where it enters the linear regime, in which we have $\partial/\partial\tau\to 0$. Thus, the transformation Eq. (\ref{rettime1}) is well behaved on the entire $\vec\rho$ plane.

In the 1-D case, $\partial A_{\bot_0}/\partial x_2 = \partial A_{\bot_0}/\partial y_2 = 0$, Eq. (\ref{eqvarphi2}) is readily solved as
$
\varphi = \int \kappa \,\, d \zeta ,
$
where $\zeta = \vec{e}_z\cdot\vec{\rho}$. Unfortunately, in 2 and 3 dimensions, Eq. (\ref{eqvarphi2}) can be solved only numerically and with considerable difficulties. It is reasonable to expect that, in a 3-D regime, $\varphi$ is of a similar order of magnitude as in 1-D and, as a consequence, our basic equations (\ref{waveqenv012}) and (\ref{poissons1}) features the following scaling
\[
2\, {\rm Re}\left\{e^{i\left[\varphi\left(t_1,\vec{r}_1\right) - t/\omega+k z\right]}\left[\alpha\,  \vec{A}_{\bot_0} -
2\, i\,\epsilon^2\, \frac{\partial \vec{A}_{\bot_0}}{\partial t_2} - 2\,i\, \epsilon\left(\nabla_\rho\varphi\cdot\nabla_2\right) \vec{A}_{\bot_0} -
\epsilon^2\,\nabla_2^2 \vec{A}_{\bot_0} + {\cal O}\left(\epsilon^3\right)
\right]\right\} =
\]
\begin{equation}
\label{waveqenv03}
\epsilon\left(\epsilon\,\frac{\partial}{\partial t_2}- u \, \frac{\partial}{\partial z_2}\right){\nabla_2}_\bot\phi -
\left(\epsilon^4\,\frac{\partial^2}{\partial t_2^2} - \epsilon^2\nabla_2^2 - \frac{1}{1-\phi}\right)\vec{A}^{(0)}_\bot,
\end{equation}
\begin{equation}\label{poissons3}
\frac{\partial^2\phi}{\partial z_2^2} = \frac{\left(\phi - 1\right)^2 - 1 - {\vec{A}_\bot}^{\, 2}}{2 \left(\phi - 1\right)^2} + {\cal O}\left(\epsilon\right).
\end{equation}

\subsection{Regime in which the self-generated magnetic field can be neglected}

As we have already mentioned, the left-hand-side of the wave equation (\ref{waveqenv012}), or (\ref{waveqenv03}), describes the evolution of the (slowly varying) envelope of the laser pulse, while the right-hand-side, whose characteristic frequency and wavevector are equal to zero, describes the slowly varying magnetic field created by the laser pulse. However, the latter is not adequately described by our simple hydrodynamic equations (\ref{bdcont})-(\ref{bdnormom}) that are based on a cold, unmagnetized plasma model. Within the accuracy with which the nonlinear current and charge densities have been derived, our wave equation (\ref{waveqenv03}) is valid only in the regime when its right-hand side is negligible. A simple scaling analysis of Eqs. (\ref{waveqenv03}) and (\ref{poissons3}) indicates that in a 3-D case ($\nabla_{2_\bot}\sim \partial/\partial z_2$)  the self-generated magnetic field (i.e. the slowly varying vector potential $\vec{A}_\bot^{(0)}$) can be neglected when
\begin{equation}\label{nomagneticfield}
{\rm max}\left(\phi, \, |\vec{A}_{\bot_0}|, \, |\vec{A}_\bot^{(0)}|\right) < \epsilon^{-1} .
\end{equation}
Particularly simple case is that of a circularly polarized wave, $\vec A_{\bot_0} = (A_{\bot_0}/\sqrt{2})(\vec e_x + \vec e_y)$, for which we readily have ${{\vec{A}_\bot}}^{\,\,2} = |A_{\bot_0}|^2$, i.e. the second harmonic is absent. Similarly, for a linearly polarized wave, we will neglect the second harmonic (because its contribution is nonresonant) and use  ${{\vec{A}_\bot}}^{\,\,2} \approx |A_{\bot_0}|^2$, where  $\vec A_{\bot_0} = A_{\bot_0}\,\vec e_x$.

Now, with the accuracy to $\epsilon^2$, our basic system of equations reduces to
\begin{equation}\label{poissons4}
\left[\alpha_{Re}\left(\phi\right) + i \, \alpha_{Im}\left(\phi\right)\right] A_{\bot_0} -
2\, i\,\epsilon^2\, \frac{\partial A_{\bot_0}}{\partial t_2} - 2\,i\, \epsilon\left(\nabla_\rho\varphi\cdot\nabla_2\right) A_{\bot_0} - \epsilon^2\,\nabla_2^2 A_{\bot_0} = 0,
\end{equation}
\begin{equation}\label{poissons04}
\frac{\partial^2\phi}{\partial z_2^2} = \frac{\left(\phi -
1\right)^2 - 1 - \left|A_{\bot_0}\right|^2}{2 \left(\phi -
1\right)^2},
\end{equation}
\begin{equation}\label{eqvarphi4}
\left(\nabla_\rho\varphi\right)^2  = \kappa^2\left(\phi\right),
\end{equation}
where the variables $(\tau, \vec\rho)$ are defined in Eq. (\ref{rettime1}) and $\alpha_{Re}$ and $\alpha_{Im}$ are the real and imaginary parts of $\alpha$, respectively
\begin{equation}\label{alfaRe}
\alpha_{Re}\left(\phi\right) = {\phi}/({1-\phi}) + \kappa^2\left(\phi\right),
\quad\quad
\alpha_{Im}\left(\phi\right) = -\nabla_1^2\varphi + \frac{\partial^2\varphi}{\partial t_1^2}= -\nabla_\rho^2\varphi +  {\cal O}(\epsilon^4). 
\end{equation}
It is worth noting that, within the adopted accuracy, the spatial integral of the intensity of the laser pulse is conserved. Multiplying Eq. (\ref{poissons4}) by $A_{\bot_0}^*$, taking the imaginary part, and integrating for the entire space, we obtain
\begin{equation}\label{conserve}
\frac{\partial }{\partial t_2}\int d^3 \vec r_2 \,\, \left|A_{\bot_0}\right|^2
=
\int d^3 \vec r_2 \, \left|A_{\bot_0}\right|^2 \left[\epsilon^{-1} \, \nabla_2 \cdot \nabla_\rho\varphi + \epsilon^{-2} \, \alpha_{Im}\left(\phi\right)\right]
= {\cal O}\left(\epsilon^2\right),
\end{equation}
which is negligible within the adopted accuracy.

\subsection{Model for the nonlinear terms $\alpha_{Re}(\phi)$ and $\kappa(\phi)$}

\begin{figure}[htb]
\centering
\includegraphics[width=140mm]{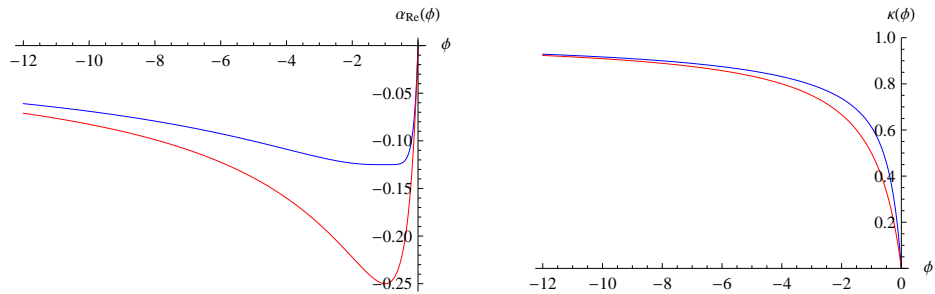}
\caption{Left: Nonlinearity functions $\alpha_{Re}(\phi) = \phi/(1 - \phi)^2 $ (red line)  and  $\alpha_{Re}(\phi) = \phi(1 + \phi)^2/(1 - \phi)^4$ (blue line).\\
Right: Functions $\kappa(\phi) = -\phi/(1 - \phi)$ (red line) and $\kappa(\phi) = [-\phi/(1 - \phi)][1+2/(1 - \phi)^2]^{1/2}$ (blue line).} \label{fig1}
\end{figure}

The simplest choice for the function $\kappa$ and the corresponding expression for $\alpha_{Re}$ [see Eq. (\ref{alfaRe})], that provide a proper asymptotic ordering of the wave equation for both $\phi\ll 1$ and $\phi\gg 1$, are given by the standard focusing-defocusing model for the {\em nonlinearity function} $\, \alpha_{Re}(\phi)$, viz.
\begin{equation}\label{choice1}
\alpha_{Re}\left(\phi\right) = \frac{\phi}{\left(1 - \phi\right)^2}
\quad \Rightarrow\quad
\kappa\left(\phi\right) = -\frac{\phi}{1 - \phi},
\end{equation}
which are displayed in red color in Fig. \ref{fig1}. In the weak and moderate intensity regimes (WIR and MIR), $|\phi|\sim\epsilon^2$, such choice of $\alpha_{Re}(\phi)$ gives the scaling $\alpha_{Re}(\phi) \sim\kappa(\phi) \sim \varphi \sim {\cal O}(\epsilon^2)$ and, keeping only the leading terms which are of the order ${\cal O}(\epsilon^2)$, our Eqs. (\ref{poissons4})-(\ref{eqvarphi4}) reduce to a Schr\"{o}dinger equation with a nonlocal cubic nonlinearity, studied in detail in our earlier paper \cite{moj_EPJD}. Conversely, in the strong intensity regime (SIR), $1 \ll \phi \lesssim 1/\epsilon$ (recall, $\phi \lesssim 1/\epsilon \,$ is the upper limit for the potential $\phi$, when we still may neglect the self-generated magnetic field), we have $\alpha_{Re}\to 1/\phi\to - \epsilon$  and  $\kappa\to 1$. Thus, in the SIR and with the accuracy to leading order in the small parameter $\epsilon$, our basic equations (\ref{poissons4})-(\ref{eqvarphi4}) reduce to a linear equation
\begin{equation}\label{poissons4SIR}
A_{\bot_0} + 2\,i\left(\nabla_\rho\varphi\cdot\nabla_2\right) A_{\bot_0} = 0,
\end{equation}
where from Eq.s (\ref{poissons04}) and (\ref{eqvarphi4}) we have $|\nabla_\rho\varphi|  = 1$  and  $\phi= - |A_{\bot_0}|$. As discussed earlier, see e.g. Ref. \cite{moj_EPJD}, the above describes a vacuum channel that has been created by a very strong laser pulse. In such a case, the ponderomotive force associated with the laser pulse is so strong that the resulting electron density perturbation, Eq. (\ref{n2}), is very large, viz. $\delta n = n-1 \sim {\cal O} (1)$, i.e. practically all electrons have been removed and the pulse propagates in vacuum. It is worth noting that the huge electric fields arising from such electron density perturbation may affect also the ions and it might be necessary to take into account also the ion dynamics (most likely in the wake).

Closer to the edges of a SIR laser pulse, in the region where $\phi\sim {\cal O}(1)$, the simple choice (\ref{choice1}) for the function $\alpha_{Re}(\phi)$ attains a relatively large value, $\alpha_{Re}(-1) = 1/4$. Thus, the nonlinear term $\alpha A_{\bot_0}$ in the wave equation becomes relatively large and it can be balanced by the linear terms only if the space/time variation of the amplitude $A_{\bot_0}(t_2,\vec{r}_2)$ is sufficiently rapid. As a consequence, in the numerical solution of the equation (\ref{poissons4})-(\ref{eqvarphi4}) for a SIR (strong intensity regime) laser pulse, there may arise numerical problems (e.g. with the convergence, stability, ...) in the spatial region where $\phi\sim {\cal O}(1)$. Such difficulty may be avoided if we adopt a different functional dependence for $\alpha_{Re}(\phi)$, which features the same asymptotic behavior for $|\phi|\ll 1$ and $|\phi| \gg 1$ as the simplest choice Eq. (\ref{choice1}), but remains sufficiently small for $|\phi| \sim 1$. Obviously, there exists a broad range of possible choices for such functions. As a simple example, we propose the following
\begin{equation}\label{choice2}
\alpha_{Re}\left(\phi\right) = \frac{\phi\left(1 + \phi^2\right)}{\left(1 - \phi\right)^4}
\quad\Rightarrow\quad
\kappa(\phi) = -\frac{\phi}{1 - \phi}\left[1+\frac{2}{\left(1 - \phi\right)^2}\right]^\frac{1}{2}.
\end{equation}
These functions are displayed in blue color in Fig. \ref{fig1}. We note that they provide the proper scaling of the wave equation both in the asymptotic regions $\phi\ll 1$ and $\phi\gg 1$, and in the intermediate domain $\phi\sim 1$.


We conclude this section by writing the wave equation in an explicit form, which is possible only in the 1-D regime.
First, we note that for $\nabla_{2_\bot}\ll\partial/\partial z_2$, the right-hand-side of the wave equation (\ref{waveqenv03}) can be neglected for arbitrarily large intensities, i.e. we are not restricted by the condition (\ref{nomagneticfield}). Next, in the 1-D regime, equation (\ref{eqvarphi2}) for the nonlinear phase can be solved as
\begin{equation}\label{WKBforphase}
{\partial\varphi}/{\partial\zeta} = \kappa\left(\phi\right) ,
\end{equation}
which after the substitution into the envelope wave equation (\ref{poissons4}) gives
\begin{equation}\label{waveq51D}
\left[\alpha_{Re}\left(\phi\right) - i \epsilon\, \frac{\partial\kappa\left(\phi\right)}{\partial z_2}\right]  A_{\bot_0} -
2\, i\,\epsilon^2\, \frac{\partial A_{\bot_0}}{\partial t_2} - 2\,i\, \epsilon\, \kappa\left(\phi\right) \frac{\partial A_{\bot_0}}{\partial z_2} - \epsilon^2\,\frac{\partial^2 A_{\bot_0}}{\partial z_2^2} = 0,
\end{equation}
where the electrostatic potential $\phi$ is found from the Poisson's equation (\ref{poissons04}), while for the functions $\kappa(\phi)$ and $\alpha_{Re}(\phi)$ we may use either the expressions (\ref{choice1}) or (\ref{choice2}).

\subsection{Madelung fluid representation and the local limit}

It is instructive to rewrite the envelope wave equation (\ref{poissons4}) within the Madelung fluid formalism, expressing the slowly varying complex amplitude $A_{\bot_0}$ via its modulus and phase, viz. $A_{\bot_0} = a \, \exp(i\, \delta\varphi)$, where $a$ and $\delta\varphi$ are purely real quantities. Obviously, this result can be obtained also directly from Eq. (\ref{waveqenv012}) with a different choice for the phase, substituting $A_{\bot_0}\to a$ and $\varphi\to\varphi' + \delta\varphi$, where $\varphi'$ is determined by $|\nabla_\rho \varphi'| = \kappa$, see Eq. (\ref{eqvarphi4}).

Separating the real and imaginary parts of the envelope equation and using the notation $\theta = \varphi + \delta\varphi$, we have
\begin{equation}\label{Madel_cont}
\frac{\partial a^2}{\partial t_2} + \nabla_2\cdot\left(a^2\nabla_2\theta\right) = 0,
\end{equation}
\begin{equation}\label{Madel_moment---}
\frac{\partial\theta}{\partial t_2} + \frac{1}{2} \left(\nabla_2\theta\right)^2 = \frac{1}{2}\left[\frac{\nabla_2^2 a}{a} - \frac{\phi}{\epsilon^2\left(1-\phi\right)}\right].
\end{equation}
Finally, taking the gradient of Eq. (\ref{Madel_moment---}), we have
\begin{equation}\label{Madel_moment}
\left(\frac{\partial}{\partial t_2} +\nabla_2\theta\cdot\nabla_2\right)\nabla_2\theta = \frac{1}{2}\, \nabla_2\left[\frac{\nabla_2^2 a}{a} - \frac{\phi}{\epsilon^2\left(1-\phi\right)}\right].
\end{equation}
Equations (\ref{Madel_cont}) and (\ref{Madel_moment}) can be interpreted as the ``continuity'' and ``momentum'' equation for the Madelung fluid, where $\nabla\theta$ and $a^2$ play the role of the fluid velocity and density, respectively. We note that for $|\phi| \gtrsim 1$, the last term on the right-hand-side of Eq. (\ref{Madel_moment---}) or (\ref{Madel_moment}) becomes very large, which makes these equations difficult to solve by standard numerical methods. However, substituting $\theta = \varphi + \delta\varphi$ and with an appropriate choice of the function $\alpha_{Re} (\phi)$, such as Eq. (\ref{choice2}), they take a convenient form in which all terms are of the same order in the small quantity $\epsilon$, viz.
\begin{equation}\label{Madel_cont_z_2}
\frac{\partial a^2}{\partial t_2} + \nabla_2 \cdot \left[a^2 \, \nabla_2\left(\varphi + \delta\varphi\right)\right] = 0,
\end{equation}
\begin{equation}\label{Madel_moment_z_2}
\left[\frac{\partial}{\partial t_2} + \nabla_2\left(\varphi + \delta\varphi\right)\cdot\nabla_2\right]\nabla_2\delta\varphi + \left(\nabla_2\delta\varphi\cdot\nabla_2\right)\nabla_2 \varphi = \frac{1}{2}\, \nabla_2\left[\frac{\nabla_2^2 a}{a} - \frac{\alpha_{Re}\left(\phi\right)}{\epsilon^2}\right],
\end{equation}
In the 1-D regime, $\partial/\partial x_2 = \partial/\partial y_2 = 0$, the Madelung hydrodynamic equations  (\ref{Madel_cont}) and (\ref{Madel_moment}) are vastly simplified if the phase $\theta$ separates variables as
$\theta(z_2,t_2) = \theta_0(t_2) + s(t_2)\, \, g(z_2)$, viz.
\begin{eqnarray}\label{Madelung1a}
&& \frac{\partial a^2}{\partial t_2} + s\left(t_2\right)\,\,\frac{\partial}{\partial z_2}\left[a^2g'\left(z_2\right)\right] = 0,
\\
&& \label{Madelung3a}
\frac{1}{a}\frac{\partial^2 a}{\partial z_2^2} - \frac{\phi}{\epsilon^2\left(1-\phi\right)} - 2\theta_0'\left(t_2\right) = 2 g\left(z_2\right)\, s'\left(t_2\right)+ {g'}^2\left(z\right) \, s^2\left(t\right).
\end{eqnarray}
The Madelung continuity equation (\ref{Madelung1a}), being a linear equation of the first order, is readily integrated as
\begin{equation}\label{Continuity_integrated}
a\left(z_2,t_2\right) = \frac{1}{\sqrt{g'\left(z_2\right)}}\,\,
F\left[\int \frac{dz_2}{g'\left(z_2\right)} - \int s\left(t_2\right)\, dt_2\right],
\end{equation}
where $F$ is an arbitrary function of its argument. The solution (\ref{Continuity_integrated}) for $a(z_2,t_2)$, must satisfy also the Madelung momentum equation (\ref{Madelung3a}) combined with the Poisson's equation (\ref{poissons04}), which can be achieved only in a limited number of cases, by a suitable choice of the functions $\theta_0(t_2)$, $s(t_2)$ and $g(z_2)$.

\begin{figure}[htb]
\includegraphics[width=80mm]{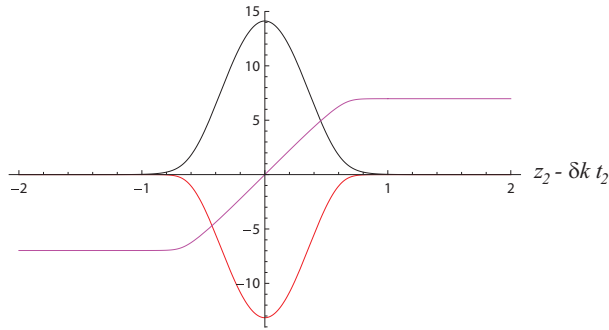}
\caption{One dimensional solution in the local Strong Intensity Regime. The laser amplitude $a_L = |A_{\bot_0}|$, given by Eq. (\ref{MadelungL7}) (black), the electrostatic potential $\phi_L = 1 - \sqrt{1 + a_L^2}$ (red), and the nonlinear phase $\varphi = \int d\zeta \,\, \kappa\left(\phi_L\right)$, where $\zeta = (z_2 - \delta k \,t_2)/\epsilon$ (magenta), are obtained adopting $C_1 = -19$, $C_2 = -2/\epsilon^2$, and $\epsilon = 1/12$. For such pulse, the full
e-folding length is $L_{\Vert ef} = 0.65$ which, in physical units, corresponds to $L_{\Vert ef} = 1.1 \, \mu{\rm m}$. (color online)}
\label{slikalocalsolution}
\end{figure}

As the simplest example, we can see that that in the local limit (denoted by the subscript "$L$"), there exists a simple 1-D localized stationary solution, whose envelope is traveling with a constant velocity. Setting $\partial^2\phi_L/ \partial z_2^2\to 0$, the Poisson's equation (\ref{poissons04}) becomes purely algebraic and it is readily solved for the electrostatic potential, as
\begin{equation}\label{local_phi}
\phi_L = 1 - \sqrt{1 + a_L^2}
.
\end{equation}
Making the simple choice $\theta_0(t_2) = -\delta\omega \,t_2$, $s(t_2) = 1$, and $g(z_2) = \delta k\,\, z_2$, where $\delta\omega$ and $\delta k$ are arbitrary constants, a linear ansatz for the phase $\theta$ is obtained, viz.
\begin{equation}\label{ansatz}
\varphi + \delta\varphi \equiv \theta\left(z_2,t_2\right) = \delta k\, z_2 - \delta \omega\, t_2 ,
\end{equation}
and the solution (\ref{Continuity_integrated}) of the ``continuity equation'' (\ref{Madel_cont}) readily yields a traveling expression for the modulus, viz. $a_L = a_L(z_2 - \delta k \,\, t_2)$, while  the ``momentum equation'' (\ref{Madel_moment}),
after a multiplication by $a_L'$, is readily integrated as
\begin{equation}\label{MadelungL5a}
{a_L'}^{\, 2}=C_2 + C_1 \, a_L^2 + \left({2}/{\epsilon^2}\right)\sqrt{1 + a_L^2}.
\end{equation}
Here the primes stand for the derivative with respect to $z_2 - \delta k \, t_2$, while $C_1$ (where $C_1\equiv\delta k^2 - 2\,\delta\omega - 1/\epsilon^2$) and $C_2$ are arbitrary constants of integration. Equation (\ref{MadelungL5a}) can be integrated by quadratures, yielding
\begin{equation}\label{MadelungL7}
z_2 - \delta k \, t_2 = \int\frac{d a_L}{\sqrt{C_2 + C_1 \, a_L^2 + \left({2}/{\epsilon^2}\right)\sqrt{1 + a_L^2}}} ,
\end{equation}
From the condition $a_L'=0$ in Eq. (\ref{MadelungL5a}), one readily finds that a solution that is vanishing at $z_2\to\pm\infty$ is obtained when $C_2 = -2/\epsilon^2$ and that its maximum value is equal to $a_{max} = (2 /C_1 \,\epsilon^2)\sqrt{1 + C_1 \, \epsilon^2}$. A typical localized solution, Eq. (\ref{MadelungL7}), with $C_1 = -19$, $C_2 = -2/\epsilon^2$, and $\epsilon = 1/12$, is displayed in Fig. \ref{slikalocalsolution}. Its maximum value $a_{max} = 14.12$ is within the {\it strong intensity regime} and the corrections due to the self-generated magnetic field are sufficiently small, see Eq. (\ref{nomagneticfield}). The e-folding length $L_{\Vert ef}$ of this solution, defined as the distance between the points where its intensity is reduced by the factor $e$, viz. $a_L^2(L_{\Vert ef}/2) = a_{max}^2/e$, is given by $L_{\Vert ef} = 0.65$, which in the physical (non-scaled) variables corresponds to $L_{\Vert ef} = 1.1 \, \mu{\rm m}$.

\section{Numerical solutions}

We have studied the influence of the nonlocality effects, arising from the derivatives in the Poisson's equation, by the numerical solution of Eqs. (\ref{poissons4})-(\ref{eqvarphi4}) and with the the function $\alpha_{Re}(\phi)$ in the form Eq. (\ref{choice2}). First, we have restricted our study to a purely 1-D regime, $\partial/\partial x_2 = \partial/\partial y_2 = 0$, in which analytical solutions are available for the nonlinear phase $\varphi$ and for the response in the local regime, see Eq.s (\ref{WKBforphase}) and (\ref{MadelungL7}) and Ref. \cite{mahajan}.

\subsection{One dimensional results}

The temporal evolution of a 1-D l laser pulse and the development of the electrostatic wake in the nonlocal Strong Intensity Regime was studied by finding the numerical solution of Eq.s (\ref{waveq51D}) and (\ref{poissons4}), using the model (\ref{choice2}) and (\ref{WKBforphase}) for the nonlinear phase $\varphi$. We used the numerical method of lines with the spatial discretization involving 900 points and 15 000 steps in time. The initial wake potential and nonlinear phase were adopted to be equal to zero, $\phi(z_2,0) = \varphi(z_2,0) = 0$, while the initial laser amplitude was taken in the form of a scaled local solution,
$
A_{\bot_0}(z_2,0) = a_L(z_2/L_z)\, \exp({i\,\delta k\, z_2}),
$
where the local solution $a_L(z_2)$ is given by Eq. (\ref{MadelungL7}) and displayed in Fig. \ref{slikalocalsolution}.

\begin{figure}[htb]
\includegraphics[width=140mm]{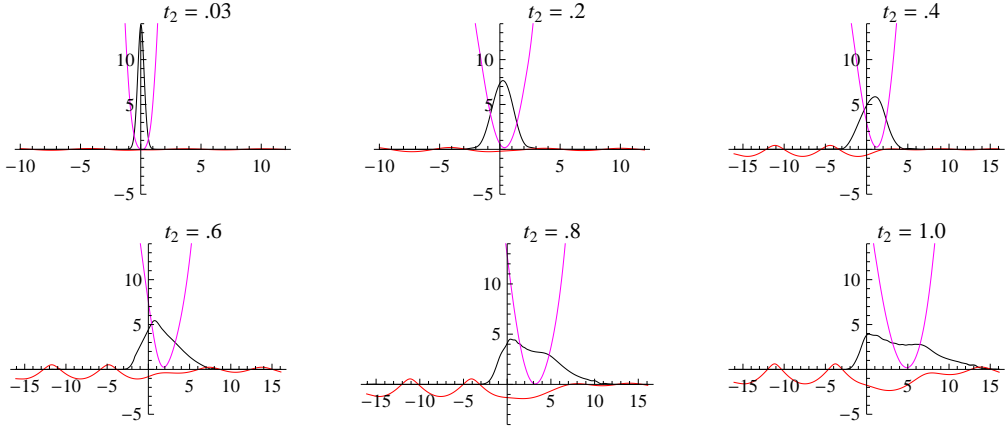}
\caption{The forward stretching of a 1-D laser pulse in the nonlocal Strong Intensity Regime  (with $\nabla_{2_\bot} A_{\bot_0}\to 0$ and $\partial^2 \phi/\partial z_2^2 \ne 0$). The laser amplitude $|A_{\bot_0}|$ (black), the electrostatic potential $\phi$ (red), and the phase $\delta\varphi = {\rm arg}\,(A_{\bot_0})$ (magenta) are obtained as the numerical solutions of Eqs. (\ref{waveq51D}) and (\ref{poissons4}), using the model (\ref{choice2}) and (\ref{WKBforphase}) for the nonlinear phase $\varphi$.  The initial condition is adopted in the form of a local solution, $A_{\bot_0}(z_2,0) = a_L(z_2/L_z)\, \exp({i\,\delta k\, z_2})$, with $\delta k = -0.5$ and $L_z = 0.8$. $a_L(z_2)$ is given by Eq. (\ref{MadelungL7}) using $C_1 = -19$, $C_2 = -2/\epsilon^2$, and $\epsilon = 1/12$, see Fig. \ref{slikalocalsolution}. (color online)} \label{sir_1-d}
\end{figure}

Even after an extensive search, we were unable to find any stationary nonlocal solutions. The longest lifetime was observed when the initial conditions were close to to the local solution, i.e. for $L_z\lesssim 1$, when the temporal evolution could be described as the breakdown of the initial pulse into two pulses propagating with the group velocity of the electromagnetic wave and with the speed of light, respectively. This disintegration was not complete since these "daughter pulses" remained attached to each other via an inclined plateau (ramp). Depending on the characteristic wavenumber $\delta k$, different patterns in the redistribution of the intensity $|A_{\bot_0}|$ among the two pulses and the plateau were observed, yielding either to the rapid spatial dispersion of the structure or its dismemberment via the generation of short scales. The longest living structure was that in which the "daughter pulses" were the least prominent, so that the  overall picture could be described as the forward stretching of the initial laser pulse into a wedge-shaped structure. As $a_L$ is independent on the characteristic wavenumber $\delta k$, the latter is a free parameter and the value $\delta k = - 0.5$ was adopted such as to maximize the lifetime of the pulse, which was determined after a large number of numerical tests. Consistent with the SIR in a pancake geometry, we adopted also $C_1 = -19$ and $L_z = 0.8$, which correspond to the initial full e-folding length of the pulse $L_{\Vert ef} = 0.52$ (in physical units $L_{\Vert ef} = 0.9 \, \mu{\rm m}$). The adopted SIR amplitude $|A_{max}| = 14.12$ (in physical units this corresponds to the wave intensity $I = 0.6 \times 10^{20} \, {\rm W/cm}^2$) is 35 times bigger than that we used previously in the MIR study, see Fig. 2 of Ref. \cite{moj_EPJD} (note that a different normalization was used).
Our initial e-folding length is $1/3$ of that used in the earlier MIR calculations \cite{moj_EPJD}.

Our 1-D SIR solution is displayed in Fig. \ref{sir_1-d}. We were able to follow its evolution until the emergence of short spatial scales at $t_{2_{max}}\gtrsim 1$ (i.e. $t_{2_{max}}\gtrsim 9.69 \times 10^{-12} \, {\rm s}$ in physical units), which practically coincides with the plasma interaction length used in the self-injection test experiment (SITE) using FLAME \cite{Gizzi2013,GizziNuovoCim}. The pulse length (which was unrealistically short at $t_2 = 0$) was rapidly stretched and increased, eventually, around tenfold. Simultaneously, the peak value of the laser amplitude rapidly dropped from $|A_{max}| \approx 14$ to $|A_{max}| \sim 5$ and remained more-less on that level. The observed temporal evolution was 3--5 times faster than in the MIR case. The electrostatic potential also developed rather rapidly, by the time $t_2 \sim 0.3$, featuring a very large first minimum with $\phi_{max} \sim -2$ and an oscillating, non-sinusoidal, wake. The electrostatic potential was around 200 times bigger than in the MIR case \cite{moj_EPJD} (note a different normalization). Due to the fast forward stretching of the laser pulse, we observed a vigorous transfiguration of the electrostatic potential $\phi$ and the emergence of a second parasitic potential minimum within the pulse. Such fast evolving potential might not be well suited for a particle acceleration.

\subsection{Two dimensional results}

The influence of the transverse effects has been studied by the numerical solution of Eqs. (\ref{poissons4})-(\ref{eqvarphi4}) and (\ref{choice2}) in the 2-D regime $\partial/\partial y_2 = 0$. The initial conditions were adopted to be the same as in the 1-D case, with $\phi\left(x_2,z_2,0\right) = \varphi\left(x_2, z_2, 0\right) = 0$, except that the laser field was taken with a Gaussian profile in the transverse direction, $A_{\bot_0}(x_2, z_2, 0) = a_L(z_2/L_z)\,\, \exp({i\,\delta k\, z_2}) \,\, \exp{(-x_2^2/2 L_x^2)}$, with $L_x = 7.5$ and, the same as before, the function $a_L$ is shown in Fig. \ref{sir_1-d} and we adopted $\delta k = -0.5$ and $L_z = 0.8$. Such laser intensity profile has the e-folding width $L_{\bot ef} = 15$, while its rms (root-mean-square) width is $L_{\bot rms} = 7.5$. In the physical (non-scaled) variables, these initial pulse length, e-folding width and rms width are approximately equal to $0.9 \, \mu{\rm m}$,  $300 \, \mu{\rm m}$,  and $150 \, \mu{\rm m}$, respectively. With the adopted $|A_{max}| \approx 14$, these correspond to the laser energy $E \approx 125 \, {\rm J}$, 18 times bigger than that of the FLAME system at present time, and the power $P\approx 50 \times 10^{15} \, {\rm W}$, which is 170 times bigger than that of FLAME.

\begin{figure}[htb]
\centering
\includegraphics[width=140mm]{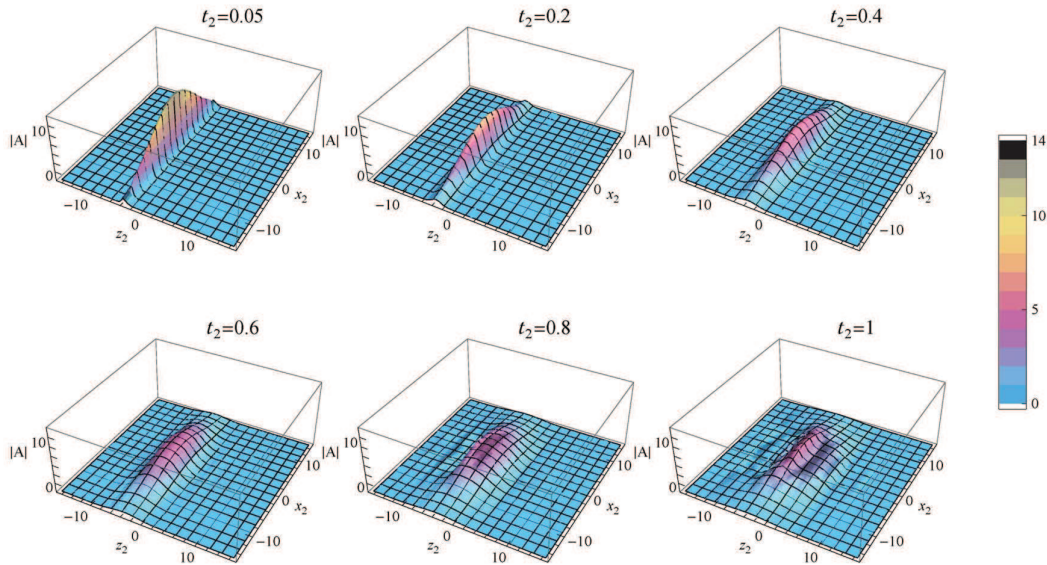}
\caption{The evolution of the envelope $|A_{\bot_0}(x_2, z_2, t_2)|$ of a laser pulse with the amplitude that is expected to be used, in the near future, for the accelerator scheme (referred to in the text as a Strong Intensity Regime). The initial condition was the same as in the 1-D case, with a Gaussian profile in the transverse direction, viz. $A_{\bot_0}(x_2, z_2, 0) = a_L(z_2/L_z)\,\, \exp({i\,\delta k\, z_2}) \,\, \exp{(-x_2^2/2 L_x^2)}$, with $\delta k = -0.5$, $L_z = 0.8$, and $L_x = 7.5$. The initial electrostatic potential and initial nonlinear phase were adopted to be zero, $\phi\left(x_2,z_2,0\right) = \varphi\left(x_2,z_2,0\right) = 0$. In the physical (non-scaled) variables, these initial pulse length and width are approximately equal to $0.9 \, \mu{\rm m}$ and  $300 \, \mu{\rm m}$, respectively. Likewise,  the dimensionless time $t_{2_{max}} = 1$ corresponds, in physical units, to $9.69\times 10^{-12} \,{\rm s}$, during which time the pulse travels $2.9 \, $mm. (color online). } \label{sir_laser}
\end{figure}

\begin{figure}[htb]
\centering
\includegraphics[width=140mm]{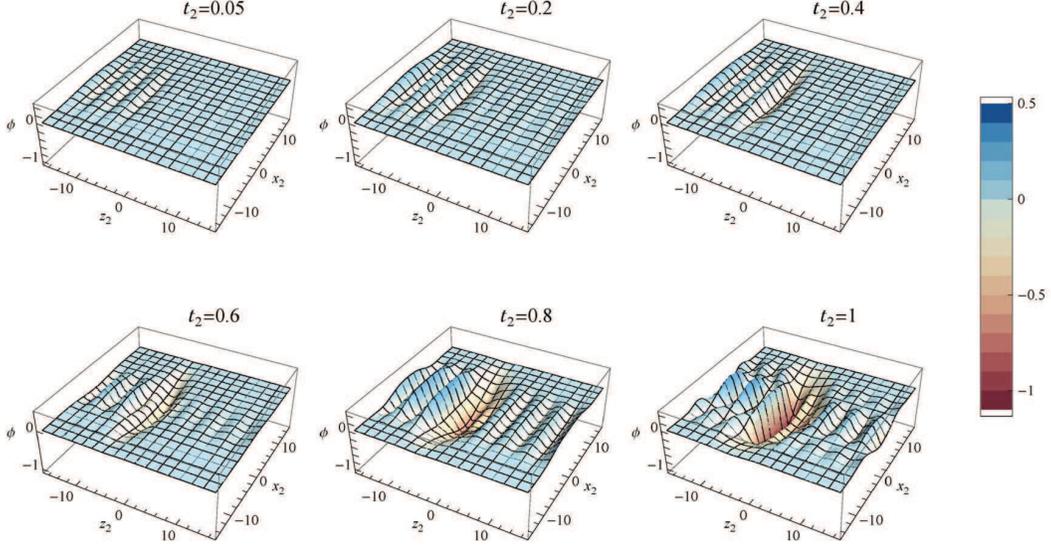}
\caption{The evolution of the electrostatic wake potential $\phi(x_2, z_2, t_2)$, produced by the laser pulse displayed in Fig. \ref{sir_laser}. A very large localized negative potential is created, with $|\phi| \gtrsim 1.5$, which indicates the almost complete expulsion of electrons in the vicinity of the laser pulse.  (color online) } \label{sir_wake}
\end{figure}

\begin{figure}[htb]
\includegraphics[width=140mm]{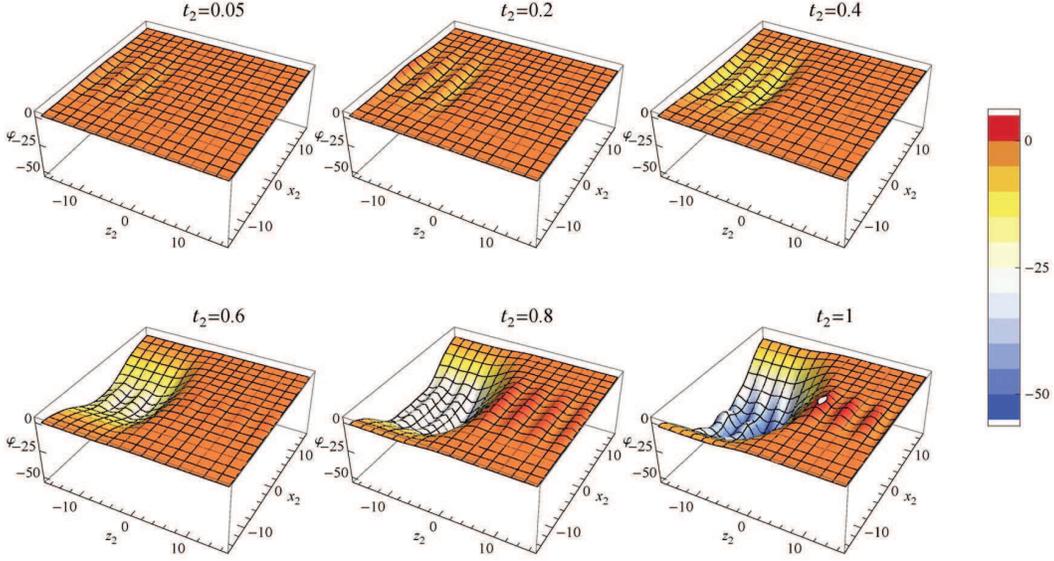}
\caption{The evolution of the nonlinear phase $\varphi(x_2, z_2, t_2)$ of the laser pulse displayed in Fig. \ref{sir_laser}. A substantial bending of the wave front occurs for $t_2 > 0.5$, simultaneously with the emergence of an electrostatic wake. (color online)} \label{sir_NLPhase}
\end{figure}

We used the numerical method of lines and an iterative procedure for Eq. (\ref{eqvarphi4}), with a discretization having $64 \times 512$ points in the $x,z$ plane and 3000 steps in time. The results are displayed in Figs. \ref{sir_laser}--\ref{sir_NLPhase}. We were was able to follow the solution up to $t_{2_{max}}\gtrsim 1$, the same as in the 1-D case. During this relatively short time no transverse collapse, or filamentation, or stretching was observed. The folding of the pancake pulse to a V-shape, known in the moderate intensity regime \cite{moj_EPJD}, was not observed, either. The initial laser amplitude rapidly dropped to $|A_{max} \sim 5$ because of a stretching in the forward direction. It was somewhat slower than in the 1-D case, and the maximum length reached by the time $t_{2_{max}}$ was around 50{\%} of that observed in 1-D. The forward stretching occurs mostly in the central part of the pulse, where the amplitude is the largest, while on the sides of the pulse it occurred both forward and backwards, and could be attributed to the linear dispersion of the wave packet. Inside the pulse, almost all electrons have been pushed out by the ponderomotive force and the laser light practically propagates in a vacuum, i.e. the core of the pulse propagates with the speed of light and the nonlinear effects in it are weak. The nonlocality produces an effective mixing with the front edge of the pulse (which tends to propagate with the group velocity), which is then carried forward by the core. The electrostatic wake develops somewhat slower than in the 1-D case, roughly during the time $t_2 = 0.5$ and it grows steadily until the maximum time is reached. The first potential minimum is quite large, but it undergoes a rather fast temporal evolution, and for this reason it might not be perfectly suited for the particle acceleration. The electrostatic potential is very large, with $|\phi_{max}| \gtrsim 1.5$. The nonlinear phase $\varphi$ emerges simultaneously with the electrostatic wake, giving rise to a substantial bending of the wave front of the laser wave.

\section{Conclusions}

In this paper we have derived nonlinear equations that appropriately describe all three intensity regimes, discussed earlier \cite{moj_EPJD}, in the interaction of an ultra-strong and ultra-short laser pulse with an unmagnetized plasma with fully relativistic electron jitter velocities. In the classical picture of a slowly varying amplitude of the laser pulse, based on a two-timescale description, it is not possible to study the strong- and moderate intensity regimes simultaneously, because the dispersion characteristics of electromagnetic waves in MIR and SIR are too different from each other and can not be described on a common footing. In the core of a very strong (i.e. SIR) pulse, the electromagnetic wave practically propagates in a vacuum. Such wave is not dispersive, i.e. its group velocity is constant and coincides with its phase velocity. Conversely, at the edges of such pulse the amplitude is smaller and the wave is dispersive. Under such conditions, the simple envelope description used previously for the MIR, breaks down.
Our novel wave equation is derived using a three-timescale description, with an intermediate timescale associated with the nonlinear phase of the electromagnetic pulse. The Schr\"{o}dinger equation for the nonlinear, amplitude-dependent phase is considerably simplified, not only under the physical conditions (the laser frequency and intensity, pulse duration and spot size, plasma density etc.) of the presently available FLAME laser system in the Frascati INFN laboratories, but also under the conditions that are envisaged with the next generation of lasers. Under those conditions, our equation for the phase can be solved within the WKB (Wentzel–Kramers–Brillouin) approximation. The new nonlinear terms in the wave equation, introduced by the nonlinear phase, describe a smooth transition to a nondispersive electromagnetic wave at very large intensities, and the simultaneous saturation of the previously known nonlocal cubic nonlinearity, without the violation of the imposed scaling laws.

In our numerical study, we solved the fully nonlinear SIR equations under the physical conditions that are consistent with the parameters used in the FLAME self-injection test experiment (SITE), but allowing for a laser power that is 170 times bigger than that used in the Plasmon-X laser wakefield experiments in Frascati. We were able to follow the solution during its travel along the distance of $2.9 \, $mm, which coincides with the dimensions of the He plasma in the the self-injection test experiment (SITE) using FLAME \cite{Gizzi2013,GizziNuovoCim}. The pulse was rapidly stretched in the forward direction, to almost ten times the initial length. Simultaneously, the peak value of the laser amplitude dropped to 1/3 of the initial value. The observed temporal evolution was 3--5 times faster than in the MIR case \cite{moj_EPJD}. The electrostatic potential also developed rather rapidly, featuring a very large first minimum and an oscillating, non-sinusoidal, wake. The potential was around 200 times bigger than that in the MIR case \cite{moj_EPJD}. Due to the fast forward stretching of the laser pulse, we observed a vigorous transfiguration of the electrostatic potential $\phi$ and the emergence of a second parasitic potential minimum within the pulse. Such fast evolving potential might not be well suited for a particle acceleration. The nonlinear phase $\varphi$ emerged simultaneously with the electrostatic wake, giving rise to a substantial bending of the laser wave front. The observed forward stretching of the laser pulse can be attributed to the combined effects of the large ponderomotive force and the nonlocality of the plasma response. In the core region, almost all electrons have been pushed out by the ponderomotive force and the laser light practically propagates in a vacuum. As a result, the core of the pulse propagates with the speed of light and the nonlinear effects in it are weak.
\begin{acknowledgements}This work was supported in part by the grant \#
171006 of the Serbian Ministry of Science and Education. One of
the authors (DJ) acknowledges financial support from the fondo FAI
of the Italian INFN and the kind hospitality of Dipartimento di
Scienze Fisiche, Universit\`{a} Federico II, Napoli.
\end{acknowledgements}



\end{document}